Manuscript published in the journal *Medical Science Monitor*
Med Sci Monit 2009; 15(9):MT121-125
Manuscript ID: 878163

# *Ex-vivo* detection of neural events using THz BioMEMS

**Abdennour Abbas[1,2], Thomas Dargent[1], Dominique Croix[2], Michel Salzet[2] and Bertrand Bocquet[1]**

[1] Institute of Electronic, Microelectronic and Nanotechnology, BioMEMS Group, UMR- CNRS 8520. University of Lille, BP 69, 59652 Villeneuve d'Ascq, France

[2] Annelids Neuroimmunology Laboratory, Leech Neuroregeneration Team, FRE CNRS 2933. University of Lille, BP 69, 59652 Villeneuve d'Ascq, France

E-mail: bertrand.bocquet@iemn.univ-lille1.fr

**Abstract**

In this paper, we present a Millimeter-wave BioMEMS (Biological MicroElectro-Mechanical System) dedicated to the *ex vivo* detection of nitric oxide synthase (NOS) activity. The latter is involved in neurodegenerative phenomena. The BioMEMS is fabricated by using polydimethylsiloxane (PDMS) sealed on a glass substrate supporting gold coplanar waveguides (CPWs). The NOS activity detection is performed by Millimeter-wave (MMW) transmission signals through CPWs placed under the lesion site of an immobilised leech nerve cord. Tests are carried out in the frequency range 140-220 GHz, and the results obtained show that MMW transmission spectroscopy combined with microfluidic technology can be used for monitoring biochemical events in aqueous environment and consequently in biological models.

## Introduction

Today, the emergence of the microtechnology, combined with the microelectronic process, allows the creation of very sophisticated miniaturized objects for biological analysis. In these integrated circuits called BioMEMS, we can mix electronic and microfluidic functions. The large majority of biosensors use the electrochemical principle or optical detection. But these techniques present some disadvantages. The principal one of them is probably the alteration of the biological activity by chemical reaction or by the fluorescent tags bonded on molecules (Facer 2001). The microwave and MMW region up to several hundred GigaHertz, could present a very interesting alternative in terms of free-label investigation (Facer *et al* 2001, Hefti *et al* 1999), and more selective detection of radicals and biomolecules (Smye *et al* 2001, Siegel *et al* 2004). This spectrum could also provide additional information in molecular biology, as for example, the conformational states of proteins (Globus *et al* 2005, Markelz *et al* 2002). The results are expressed in terms of dielectric spectroscopy. Alternatively, an approach in far field is less attractive due to the poor spatial resolution of the wavelength. The use of planar waveguides inside integrated circuits makes possible these measurements. Nevertheless, increasing the frequency range up to the TeraHertz spectrum potentially has several interesting advantages.

Here, we want to detect the nitric oxide synthase (NOS) activity in the leech nerve cord after injury. This study is of great interest due to the critical role of the nitric oxide (NO), a gaseous molecule produced by NOS activity, in several biochemical networks, especially in the enhancement of nerve cord repair (Chen *et al* 2000). Today, the detection of NOS activity is performed by histochemical staining that requires fluorescent tags, whereas the monitoring of NO generation is generally realized by amperometric measurements (Amatore 2006). But these techniques remain difficult and depend strongly on the measurement conditions, such as the distance between the probe and the nerve cord, the molecular diffusion phenomena and the short half-life of NO in aqueous media (~ 6 s). The idea here is to immobilize a nerve cord inside a MMW microfluidic system (Fig. 1(a)). We have validated this approach by using a mixed technology such as polymer on silicon for the analysis of protein solutions (Mille *et al* 2006) or living cells (Treizebré *et al* 2005, 2008). The present work is dedicated to neural tissues investigations.

## Experimental

### *Animal Model*

Our biological model is the leech *Hirudo medicinalis*, which is a very interesting and largely studied invertebrate. A lesion or a cut of its nerve cord, that represents the central nervous system (CNS), causes biochemical events leading to a complete neuron regeneration and restoration of the biological function in approximatively four weeks after damage (Modney *et al* 1997). Note that in mammals, the synapse regeneration is successful in the peripheral nervous system (PNS), but fail very quickly in the CNS (Fawcett 1997). After nerve cord injury, the fast production of NO is one of the early events occurring at the lesion (Kumar *et al* 2001). This production is catalysed by an isoenzymes family called: NO synthases (NOS, EC. 1.14.13.39), according to the reaction (1):

$$\text{L-arginine} + 2\ \text{NADPH} + 2\ O_2 + 2\ H^+ \rightarrow \text{NO} + \text{L-citrulline} + 2\ \text{NADP}^+ + 2\ H_2O \qquad (1)$$

(NADPH: reduced form of Nicotinamide Adenine Dinucleotide Phosphate)

Moreover, a treatment with the NOS inhibitor L-$N^G$-Nitroarginine methyl ester (L-NAME) blocks the repair mechanism, showing the important role of the NOS to modulate the axon growth (Chen *et al* 2000, Shafer *et al* 1998). However, the application of exogenous NO with the NO donor Spermine NONOate (SPNO) increases the concentration and leads to a total blocking of regeneration. In mammals, under pathological circumstances, the important increase in NOS activity induces toxicity and/or neurons apoptosis (Bonfoco *et al* 1995). The produced NO is involved in neurodegenerative phenomena, particularly Alzheimer's diseases and Parkinsonism (Zhang *et al* 2006). An important question is: how the precise regulation of the NOS activity in the leech induces neuron regeneration and not cell death? The real time monitoring of NOS activity is a main step towards understanding the role of NO in nerve repair and how man loses his capacity to regenerate his CNS during the neurodegenerative disorders.

*BioMEMS Design*

Before designing our microsystem, the average dimensions of the connectives and ganglions of the leech *Hirudo medicinalis* were determined. The microchannel (width: 450 µm, depth: 1 mm, length: 4 cm, volume less than 15 µl) is matched to the nerve cord dimensions for improving its immobilization and limiting the medium evaporation. The MMW BioMEMS design uses a classical technology based on PDMS due to the large dimensions of the microchannel (McDonald *et al* 2000). The PDMS is prepared by mixing the precursor Sylgard 184 (*Dow Corning*) with the crosslinking agent at 10:1 weight ratio. We use a coplanar waveguide (CPW) which is well suited to this measurement (Gupta *et al* 1996). It is placed under the probes of the vectorial network analyzer (VNA) (Fig. 1(b)).

This planar waveguide is constituted by three conductors and designed on a glass substrate (Fig. 2(a)). The latter has low losses in the MMW propagation and its transparency allows further use of contrast phase microscopy for the observation of the nerve cord under measurement. We have optimized the propagation condition on the real microsystem composed by a tank filled with liquid water. Many geometric configurations in terms of slot width to central conductor width ratio are possible in order to obtain the wanted impedance of 50 Ω, which is recommended for the connection at the VNA. Simulations are performed with a full 3D software named Microwave Studio® from Computer Simulation Technology (CST) and based on a finite element method (FEM) model. The true sizes take into account the propagation of the fundamental non-dispersive mode and a small excursion of the electric field for reducing the radiated losses. An example of a simulated structure is shown on the Figure 2(b). We obtain a good correlation between the computed and the measured values on water (Fig. 3).

The microsystem is realized on a 2-inch-square-glass substrate. A thin binding of 200 Å Chromium (Cr) layer was sputtered on the substrate before the deposition of a 0.5 µm Gold (Au) layer. Then, both Gold and Chromium are etched in order to create the CPW lines. Two-step photolithography has been adopted for the substrate processing. The PDMS channel was obtained by molding in a mechanically etched piece of Teflon®. To obtain a permanent bond of the device, both glass substrate and PDMS channel were exposed separately to plasma oxidation before joining them together.

*Spectral Measurements*

The nervous chains were extracted from adult leeches (*Hirudo medicinalis*) weighing from 2 to 3 mg, and maintained in Ringer's solution (pH 7.4) for their survival. They are manually placed and immobilized inside the open microchannel after PDMS-glass bonding. Dissection and lesions performed during the experiment are carried out in aseptic conditions using Patscheff scissors. L-NAME hydrochloride was purchased from *Cayman Chemical*. Solutions with different concentrations were prepared by dissolving L-NAME powder in the Ringer's solution, and injections were performed using a *Hamilton* Syringe, gauge 33 (internal diameter: 0.21 mm) obtained at *NHamilton Bio*. The transmission measurements were carried out at room temperature, in aerobic conditions. The VNA is an Anritsu 37147C associated with mixers of reference V05VNA2-T/R from OML (*Oleson Microwave Laboratories*) working in a bandwidth of 140-220 GHz. We use Line-Reflect-Match (LRM) calibration with a calibration kit reference 101-190B of Cascade Microtech. The analyzer is gauged in the plane of the measurement probes. Here, we have exploited only the transmission modulus relative to the wave absorption.

## Results and Discussion

In order to study the transmission spectral characteristics of the CPW, we initially carried out tests separately on well-known deionized water and on the Ringer's solution (115 mM NaCl, 4 mM $KCl_2$, and 1.8 mM $CaCl_2$), buffered at pH 7.4 with 10 mM Tris maleate (Fig. 3(a)). As it can be seen with the computed values obtained on pure water, the transmission coefficient decreases with increasing frequency, but remains valuable for a good measurement. The very small volume matches our measurements well despite the high water absorption. Note that we can observe some small rebound phenomena due to the standing waves minimized by the impedance optimisation. The same figure shows a very small difference between the spectral responses of water and Ringer's solution, which can be explained by the fact that moderate ionic concentrations have negligible effects on transmission spectra (Xu *et al* 2007). The measured values were stables and reproducibles and the measurement errors are estimated at ±0.1 dB. This result shows that the Ringer's solution can be a useful survey medium for *in vivo*/*ex vivo* studies by MMW spectroscopy. Therefore, when the nerve cord is immobilized inside the microchannel containing Ringer's solution, we observe an improvement of the transmission coefficient (Fig. 3(a) and (b)). This is mainly due to the displacement of water by nerve tissues.

The second step was the calibration with L-NAME in the millimolar concentration range. Figure 4 shows that the Beer's law is validated for the millimolar concentration, in the frequency range 145-200 GHz. We can see clearly a linear change in MMW transmission with the concentration of L-NAME. The increase in transmission in this case could be explained by the increasing number of bound water molecules (hydration shell) that present low-mobility and consequently lower dielectric constant than free water (Mickana *et al* 2002). In these experiments, the detection limit of the bioMEMS reaches down to 0.01 g/ml. This limit, obtained by varying L-NAME concentrations, concerns only hydrated biomolecules that increase transmission and not the NOS activity products (probably NO) that increase absorption as we can see below. The last step of these experiments is a preliminary comparative study of the leech nerve cord before and after injury. The aim is to block the NOS activity after injury, and observe the change in MMW transmission features. First, nerve cords were injured directly inside the microchannel filled with Ringer's solution without any treatment (Fig. 5). As it can be seen, the lesion causes a decreased transmission coefficient from approximately 0.8±0.1 dB comparatively with the intact cord. Curves (b) and (c) show respectively that after injury, NOS activity reaches immediately a peak of intensity, and decrease over time never returning to preinjury level. Secondly, nerve cords were exposed to 1 mM L-NAME during 40 min before injury. L-NAME is the antagonist of L-arginine amino-acid. Thus, it inhibits the NO synthase enzyme and then block the NO and L-citruline production. Here, the L-NAME was used as a reference sample, and any statistically significant differences in the transmission of the intact nerve cord and the nerve cord injured was observed after L-NAME treatment. Similar results obtained by the amperometric method are described in the literature (Kumar *et al* 2001). Note that the L-NAME concentration at 1 mM have no effect on transmission features according to the limited sensitivity of the present bioMEMS. The transmission change described above is likely due to the highest absorbance of millimeter-waves by the released products.

**Conclusions**

Studies of the interaction between millimeter-waves and reactive molecules in liquid phase would lead to a better understanding of the biological activity. This work demonstrates that the combination between millimeter-waves and microfluidic circuits inside a BioMEMS allows us to perform *ex vivo* measurements. The main interest is the integration on a micrometric scale to obtain fine spatial resolution and selective detection. The second advantage is the very small volume which lead to a huge increase in the local concentration of the reaction products. We have specifically designed a microsystem for *ex vivo* analysis of leech nerve cords and the preliminary results show two main considerations: Firstly, our measurements show clearly that it is possible to use millimeter wave spectroscopy for the detection of biochemical reactions in aqueous solution. Secondly, we have characterized the NOS activity around a leech nerve cord. Future experimental tests will be dedicated to the investigation of analytical parameters and the real time monitoring of NOS activity. We will obtain quantitative measurements of the reaction products for studying the effects of several drugs on NO production, and then on nerve repair mechanisms.

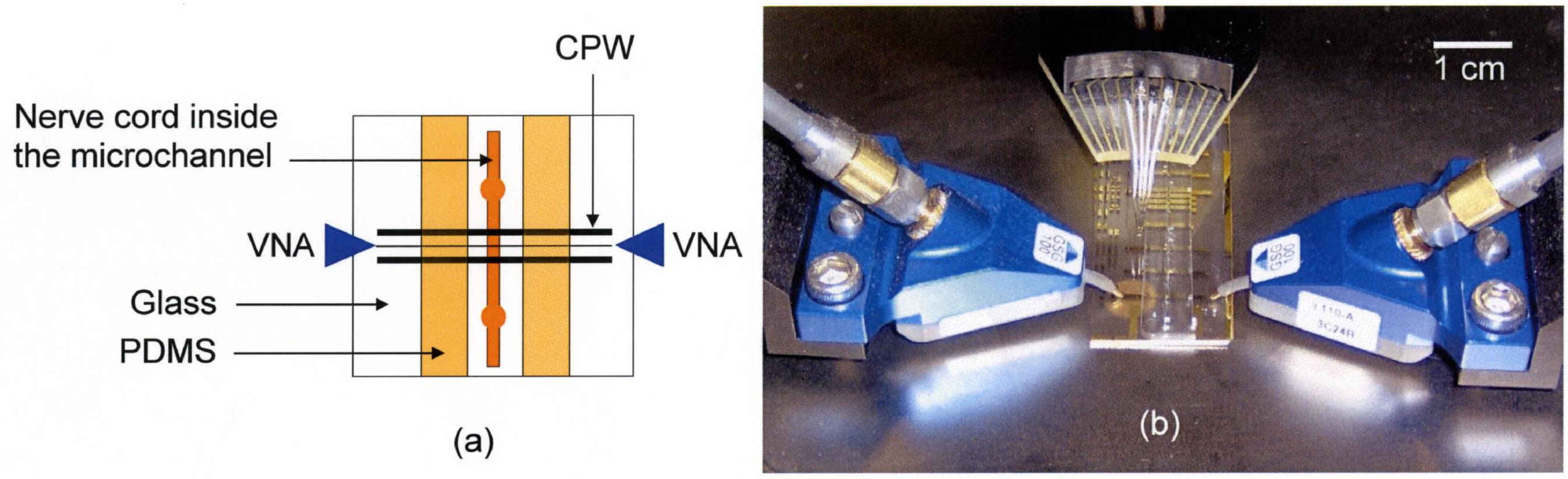

**Figure 1**

**Figure 1.** (a) Top view of immobilized nerve cord in the microchannel probed orthogonally by the VNA. (b) Photo of the bioMEMS (rectangular transparent structure) positioned between two VNA probes (in blue).

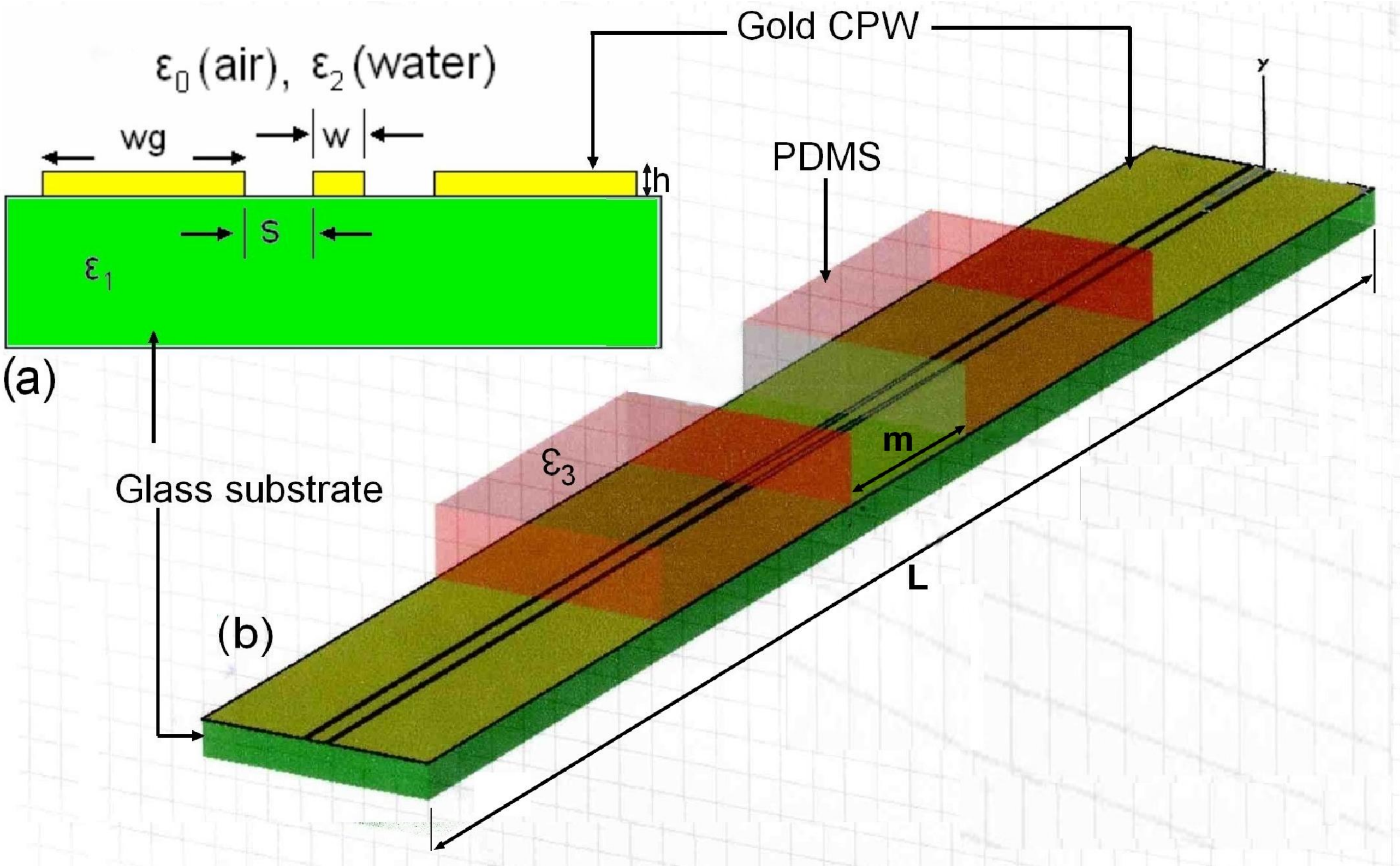

**Figure 2**

**Figure 2.** Simulated structure. (a) Cross section of the CPW: wg= 300 µm, w= 48 µm, s= 10 µm, h= 0.5 µm, $\varepsilon_0$= 1, $\varepsilon_1$= 4.82 with tan $\delta_1$= 0.0054, $\varepsilon_2$ is defined by the Debye 2$^{nd}$ order model (ε static= 77.97, $\varepsilon_\infty$= 4.59, relaxation time = 8.32 ps). (b) 3D view of cross section of the microsystem with two walls of 3 mm PDMS, m (microchannel width) = 450 µm, L (CPW length) =11 mm, $\varepsilon_3$= 2.68 with tan $\delta_3$= 0.11.

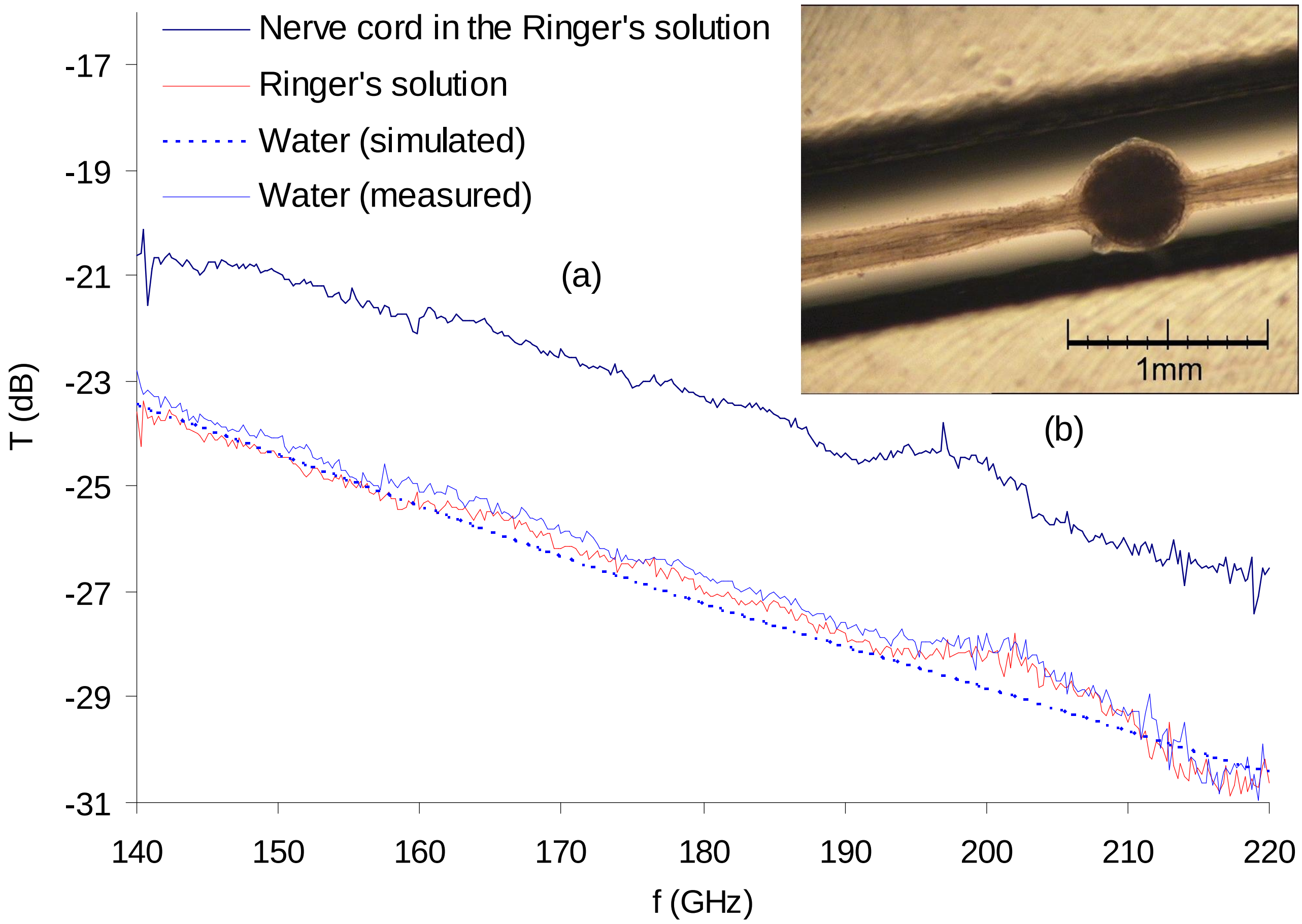

**Figure 3**

**Figure 3.** (a) Transmission curves of deionized water, Ringer's solution and nerve cord in the Ringer's solution. Simulated (blue dashed line) and experimental (blue solid line) transmission of the CPW on the water show a good correlation between the measured and the computed values. (b) The nerve cord composed of ganglions (dark disk) and connectives is lightly stretched inside the PDMS microchannel.

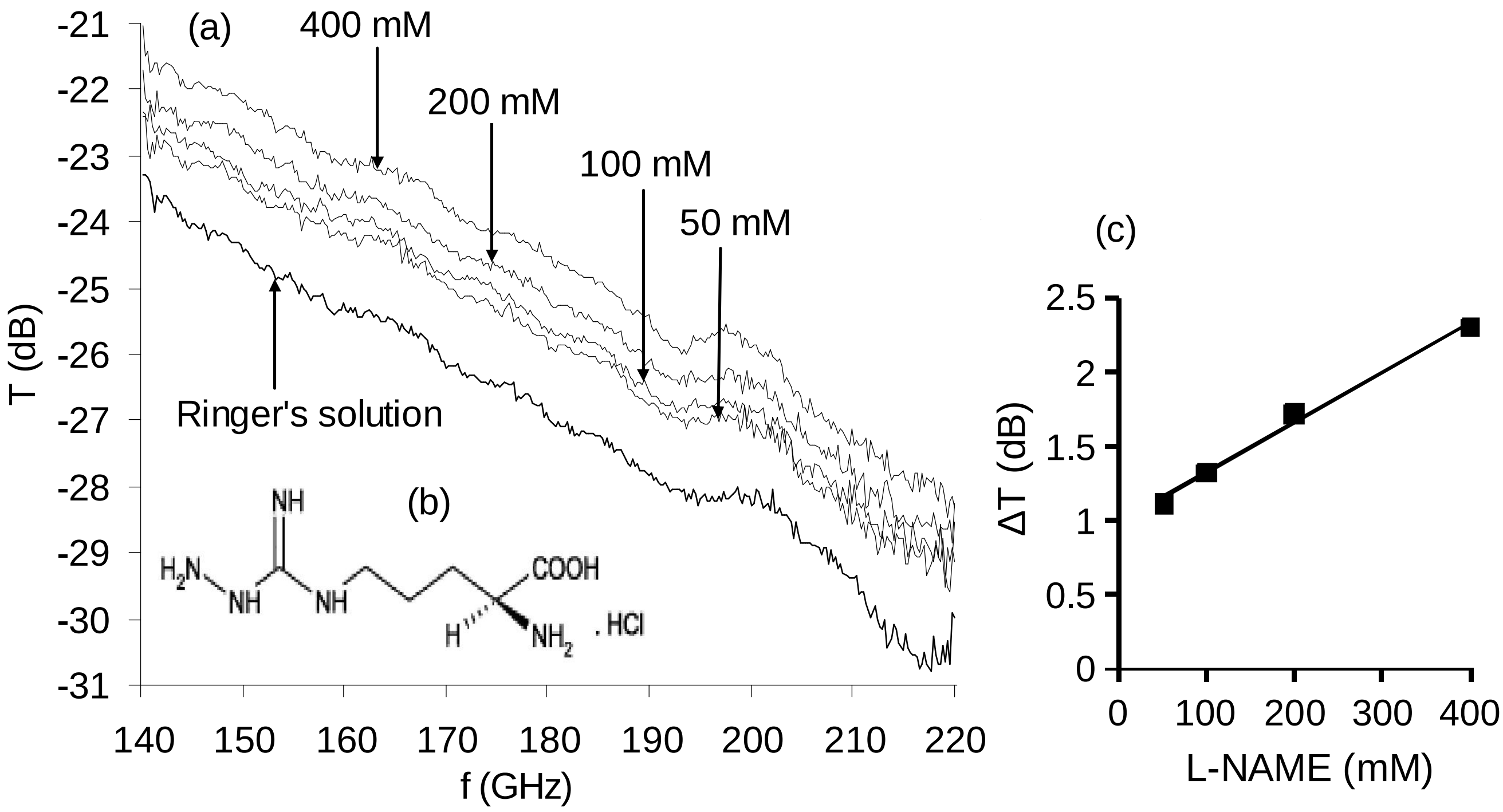

**Figure 4**

**Figure 4**. (a) Transmission curves obtained at various concentrations of solvated L-NAME in the Ringer's solution. (b) Semi-developed chemical formula of L-NAME. (c) Linear change in MMW transmission with the concentration of L-NAME. ΔT is obtained by the average values of the substraction of the waveguide transmission in the presence of Ringer's solution ($\Delta T = T_{total} - T_{Ringer}$) for each frequency in the range 145–200 GHz.

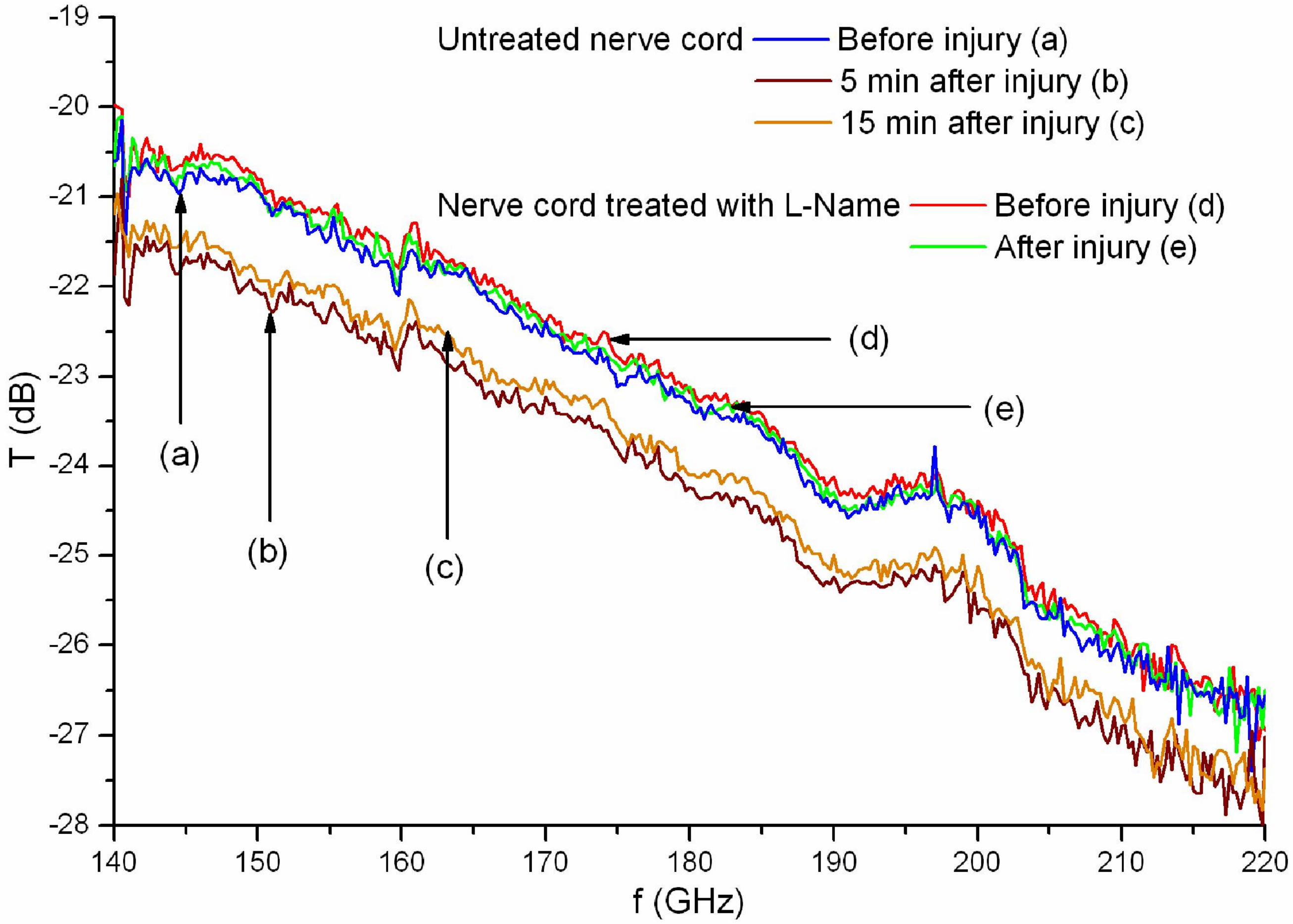

**Figure 5**

**Figure 5.** Transmission curves obtained with nerve cords in different conditions in the Ringer's solution. An injury of the nerve cord producing NO and L-citruline creates a significative change of 0.8 dB (b) and (c) compared to the reference signal obtained with intact cord (a). The reproducibility of these measurements is validated on six different cords. L-NAME is used here to block the NO production after injury. There is no significative change in transmission curves of nerve cords treated with L-NAME before (d) and after injury (e), indicating that change in MMW transmission is due mainly to NO and/or L-citruline released by the nerve cord after the lesion.